\documentclass{aa}
\usepackage{graphicx}
\usepackage{txfonts}
\begin{document} 
   \title{The stellar content of OMC\,2/3
   \thanks{Based on observations collected at the European Southern
          Observatory, La Silla, Chile}}

   \author{M. Nielbock\inst{1,2}
           \and
           R. Chini\inst{1}
           \and
           S. A. H. M\"uller\inst{1}
          }
   \offprints{M. Nielbock}

\institute{
  Astronomisches Institut der Ruhr-Universit\"at Bochum,
  Universit\"atsstrasse 159/NA 7, 44780 Bochum, Germany
  \and
  SEST, European Southern Observatory, Alonso de Cordova 3107, Santiago,
  Chile}

   \date{Received ; accepted }

   \authorrunning{M. Nielbock et al.}

\abstract{We have revisited the stellar content of OMC\,2 and OMC\,3
by means of MIR imaging and NIR photometry; in addition, we have
extended the existing (sub)mm maps by a huge $1200\,\mu$m map obtained with
SIMBA showing new sources and filamentary features for the first
time at that wavelength. The MIR data reveal 43 new sources at $N$ and
$Q$ which are partly associated with dense condensations at millimetre
wavelengths. Six close binary sources could be resolved at locations
where existing (sub)mm maps only show single emission peaks; three of
them are classified as early (B-type) systems, one of them is
compatible with type K while the remaining two seem to be T\,Tauri stars.
Furthermore, the MIR images indicate the existence of separate circumstellar
discs in the K-binary system. NIR colour--colour and colour--magnitude
diagrams obtained from 2MASS data are examined to explore the physical
properties of the sources and to derive the distribution of $J$
luminosities. There is a clear decrease in luminosity and thus in
stellar mass when going from South to North. Likewise, there is an
anti-correlation between Class\,I and II objects in both regions:
while OMC\,2 contains twice as much Class\,II objects compared to
Class\,I, the situation is reversed in OMC\,3. 
}
\maketitle
%
%
\section{Introduction}

The GMC complex Orion\,A is usually separated into three distinct
components, i.e. OMC\,1, 2 and 3. The OMC\,2/3 region has turned out to
be one of the most active sites of ongoing low and intermediate-mass
star formation known today.  Numerous observations at optical, NIR,
FIR, sub-mm and cm wavelengths have been devoted to the stellar content
of this region, and have revealed a young association with sources of
different evolutionary stages.

Gatley et al.~(\cite{gatley}) originally discovered OMC\,2 as a star
forming region by IR and CO measurements. They found 5 embedded IR
sources within a cloud core of $\sim 1000\,M_\odot$. This region is a
small fraction of the cloud complex Orion\,A found by Kutner et
al.~(\cite{kutner}). A spatially well resolved study in the $^{13}$CO
line revealed a filamentary structure (Bally et al.~\cite{bally}) of
the entire region.  Infrared studies (e.g. Rayner et
al.~\cite{rayner}; Johnson et al.~\cite{johnson}; Jones et
al.~\cite{jones}; Ali \& DePoy~\cite{ali}) discovered a population of
young stars, many of them surrounded by circumstellar discs
(e.g. O'Dell et al.~\cite{odell}; McCaughrean \& O'Dell~\cite{mc};
Hillenbrand et al.~\cite{hillenbrand}).

During the first $1300\,\mu$m continuum mapping of OMC\,1 and 2, Mezger
et al.~(\cite{mezger}) found 10 compact sources. Subsequently, Chini
et al.~(\cite{chini}) investigated OMC\,2 and 3 at $1300\,\mu$m with a
better spatial coverage and higher sensitivity, and discovered a long
integral shaped narrow dust ridge ranging from OMC\,1 to OMC\,3 that
contains 21 compact sources. Two of them were also detected by {\em
IRAS}. 350\,$\mu$m imaging by Lis et al.~(\cite{lis}) suggested the
presence of even 33 dust condensations within the area.
Finally, the largest $1200\,\mu$m continuum map of Orion\,A
(Nyman et al.~\cite{nyman}) extends over $1^\circ$ in declination and
depicts the total dust emission from that region. Many of the (sub)mm
objects exhibit molecular H$_{\rm 2}$ outflows (Castets \&
Langer~\cite{castets}; Yu et al.~\cite{yu}).

A VLA survey by Reipurth et al.~(\cite{reipurth}) led to the detection
of a number of radio sources, 11 of which were associated with
previously detected (sub)mm condensations. Tsuboi et
al.~(\cite{tsuboi01})  report on X-ray sources in OMC\,3 and associate
them with very young stellar objects.

\begin{table*}[ht]
\caption{\label{pos.tab} Source positions and other designations of
the new MIR sources. The braced brackets indicate those double sources
which were not resolved by the millimetre and/or the NIR
observations. The millimetre detections are marked in the column
``MM'' according to Chini et al.~(\cite{chini}); those millimetre
peaks without a previous designation are marked as ``+''; ``S(c)''
indicates the ``c'' component of the the new mm detection with
SIMBA. Two identifications (FIR\,4 and 5) are uncertain and thus
marked as "(?)". The sources listed in the column ``CSO'' are taken
from Lis et al.~(\cite{lis}) and the column ``VLA'' contains
detections published by Reipurth et al.~(\cite{reipurth}).  Coinciding
NIR and FIR sources are listed in the next five columns.  The
``2MASS'' numbers are from 2MASS scan No. 10; ``AD'' and ``CHS''
numbers are taken from Ali \& DePoy~(\cite{ali}) and Carpenter et
al.~(\cite{carpenter}). The ``IRS'' sources were published in Gatley
et al.~(\cite{gatley});  {\em IRAS} sources, whose positions match
within their error ellipses, are also mentioned. The line between
the sources MIR~12 and 13 indicates the borderline between OMC~2 and 3.}
\tabcolsep8.78pt
\begin{tabular}{rrrlcccccccc}
\hline
\hline
    & \multicolumn{1}{c}{RA}
    & \multicolumn{1}{c}{Dec}
    &
    & \multicolumn{8}{c}{other designations} \\
MIR & \multicolumn{2}{c}{(J2000)}& & MM & CSO & VLA & 2MASS & AD & CHS
    & IRS & IRAS\\
\hline
 1 & 5$^{\rm h}$35$^{\rm m}$28\fs29 & $-$4$^\circ$58$'$37\farcs5
                              &  &   &   &   &  & &  & &  \\
 2 & 28\fs33 & 58$'$38\farcs5 &\raisebox{4pt}[0pt]{\LARGE$\rbrace$}
                              &\raisebox{5pt}[0pt]{S(c)}
                              &
                              & 
                              &\raisebox{6pt}[0pt]{2083}
                              &
                              &\raisebox{6pt}[0pt]{10398} &  & \\
 3 & 16\fs17 & $-$5$^\circ$00$'$02\farcs6
                              &  & + & 3 &   & 2068 & & 8787 & & \\
 4 & 17\fs74 & 00$'$31\farcs9 &  &   &   &   & 2063 & &      & & \\
 5 & 18\fs34 & 00$'$32\farcs8 &  &   &   & 1 &      & &      & & \\
 6 & 18\fs33 & 00$'$34\farcs0 &\raisebox{4pt}[0pt]{\LARGE$\rbrace$}
                              &\raisebox{5pt}[0pt]{MMS\,2}
                              &\raisebox{5pt}[0pt]{6} 
                              &  
                              &\raisebox{5pt}[0pt]{2062} & & & & \\
 7 & 19\fs98 & 01$'$02\farcs9 & & MMS\,4 & 8 &  & 2053 & & & & \\
 8 & 23\fs6\hspace*{4.5pt}& 01$'$46$''$\hspace*{4.5pt}
                              &  &   &   &   &      & &  & & \\
 9 & 28\fs31 & 03$'$40\farcs8 &  & + & 11 &  & 2014 & &  & & \\
10 & 26\fs69 & 03$'$54\farcs9 & & MMS\,7 & 12 & 4 
                              & 2006 & 2117 & & & 05329-0505 \\
11 & 31\fs54 & 05$'$47\farcs3 & & + &  &  & 1979 & 2468 & 10792 & & \\
12 & 25\fs76 & 05$'$57\farcs9 & & + &  &  & 1975 & 2079 &       & & \\
\hline
13 & 21\fs89 & 07$'$01\farcs8 & &  &  &  
                              & 1957 & & 9617 & & 05329-0508 \\
14 & 23\fs34 & 07$'$09\farcs8 & & FIR\,1c & 16 &  & 1954 & 2066 &  & & \\
15 & 22\fs37 & 07$'$39\farcs1 & &  &  &   & 1948 & 2064 &      &   & \\
16 & 25\fs73 & 07$'$46\farcs2 & &  &  &   & 1946 &      & 10104 & & \\
17 & 25\fs58 & 07$'$57\farcs4 & &  &  & 9 & 1936 & 2062 &       & & \\
18 & 23\fs34 & 08$'$21\farcs6 & &  &    &  & 1925 & 1928 &       & &  \\
19 & 24\fs30 & 08$'$31\farcs0 & & FIR\,2 & 20 &  &    &      &       & &  \\
20 & 26\fs86 & 09$'$24\farcs4 & & + &  &  & 1896 &      & 10242 & 2 & \\
21 & 27\fs65 & 09$'$34\farcs0 & &  &  & 11 &     & 2446 &       & 4-N &\\
22 & 27\fs65 & 09$'$37\farcs2 &\raisebox{4pt}[0pt]{\LARGE$\rbrace$}
                              &\raisebox{5pt}[0pt]{FIR\,3}
                              &\raisebox{5pt}[0pt]{22} &  
                              & 1886 & & 10318 & 4-S &\\  
23 & 27\fs48 & 09$'$44\farcs2 & &  &  &  & 1882 & 2445 & & & \\
24 & 26\fs99 & 09$'$54\farcs5 & & FIR\,4(?) & 23 & 12 & 1876 & 1867 & & & \\
25 & 27\fs56 & 10$'$08\farcs5 & &  +        &    &    & 1868 & 1863 & & & \\
26 & 28\fs16 & 10$'$13\farcs9 & &  +        &    &    & 1864 & 2996 & 10383& & \\ 
27 & 26\fs98 & 10$'$17\farcs3 & & FIR\,5(?) & 24 &    
                              & 1862 & 2444 & & 3 & 05329-0512 \\
28 & 24\fs76 & 10$'$29\farcs6 & & + &  & 13 & 1854 & 2443 & & 1 & \\
29 & 23\fs37 & 12$'$03\farcs0 & & FIR\,6b & 25 &    & 1822 & 1691 & & & \\
30 & 18\fs23 & 13$'$06\farcs9 & & + &  &    & & 2440 & 9094 & & \\
31 & 20\fs14 & 13$'$13\farcs2 & &   &  &    & & 2405 &      & & \\
32 & 20\fs16 & 13$'$15\farcs6 &\raisebox{4pt}[0pt]{\LARGE$\rbrace$}
                              &\raisebox{5pt}[0pt]{FIR\,6d}
                              &\raisebox{5pt}[0pt]{30} &  
                              & 1796 & 2439 & 9373 & & \\
33 & 19\fs67 & 13$'$26\farcs5 & &  &  &  & 1793 & 1679 & 9305 & & \\
34 & 18\fs53 & 13$'$38\farcs7 & &  &  &  &      & 2404 & 9147 & & \\
35 & 20\fs25 & 13$'$59\farcs4 & &  &  &  & 1775 & 2403 &      & & \\
36 & 22\fs63 & 14$'$11\farcs3 & & + &  &  & 1770 & 1667 &     & & \\
37 & 21\fs94 & 14$'$27\farcs6 & & + & 31 &  & 1762 &    &     & & \\
38 & 21\fs92 & 15$'$01\farcs1 & & + & 32 &  & 1748 & 1654 & 9624 & & \\
39 & 19\fs85 & 15$'$08\farcs9 & &  &  &  &  & & & & \\
40 & 19\fs81 & 15$'$09\farcs5 &\raisebox{4pt}[0pt]{\LARGE$\rbrace$}
                              &\raisebox{5pt}[0pt]{+} &  &  
                              &\raisebox{5pt}[0pt]{1743}
                              &\raisebox{5pt}[0pt]{2402}
                              &\raisebox{5pt}[0pt]{9335} & & \\
41 & 23\fs49 & 15$'$23\farcs3 & & + &  &  & 1736 & 1646 & 9829 & & \\
42 & 19\fs83 & 15$'$35\farcs2 & & + & 33 &  & 1731 & 1644 & 9333 & & \\
43 & 25\fs30 & 15$'$35\farcs5 & &  &  &  &  & & & & \\
44 & 25\fs24 & 15$'$35\farcs7 &\raisebox{4pt}[0pt]{\LARGE$\rbrace$}
                              &\raisebox{5pt}[0pt]{+} &  &  
                              &\raisebox{5pt}[0pt]{1730}
                              &\raisebox{5pt}[0pt]{2401}
                              &\raisebox{5pt}[0pt]{10037} & & \\
45 & 20\fs76 & 15$'$49\farcs2 & &  &  &  & 1722 & 2994 & & & \\
\hline
\end{tabular}
\end{table*}

\begin{table*}
\caption[]{\label{phot.tab} Photometric data and spectral properties.
The MIR 1$\sigma$ noise level of one image from our TIMMI\,2 measurements is
typically 25\,mJy/$\sq\arcsec$ in the $N$-band and about
370\,mJy/$\sq\arcsec$ in the $Q$-band. The $350\,\mu$m data come from
Lis et al.~(\cite{lis}), the $1300\,\mu$m flux densities were
determined from the extended map of Chini et al.~(\cite{chini}), and
have a 1$\sigma$ noise level of 15\,mJy/beam. Newly determined
millimetre fluxes are marked in bold letters. VLA data at 3.6\,cm are
taken from Reipurth et al.~(\cite{reipurth}).}
\begin{tabular}{|r|rrrrrrrrr|rc|}
\hline
    & \multicolumn{9}{c|}{$S$ [mJy]} & &  \\
MIR & \multicolumn{1}{c}{$J$}
    & \multicolumn{1}{c}{$H$}
    & \multicolumn{1}{c}{$K_s$}
    & \multicolumn{1}{c}{$10.4~\mu$m}
    & \multicolumn{1}{c}{$11.9~\mu$m}
    & \multicolumn{1}{c}{$17.8~\mu$m}
    & \multicolumn{1}{c}{$350~\mu$m}
    & \multicolumn{1}{c}{$1300~\mu$m} 
    & \multicolumn{1}{c|}{3.6~cm}
    & \multicolumn{1}{r}{$\alpha_{KN}$}
    & \multicolumn{1}{c|}{$L_{\rm bol}/L_{\rm smm}$} \\
\hline
 1 &     &     &      &   30 &  &   &   & & &    &      \\
 2 & \raisebox{6pt}[0pt]{$< 0.2$}
   & \raisebox{6pt}[0pt]{2.3}
   & \raisebox{6pt}[0pt]{18.5}
   & 30
   &
   &   
   &
   & \raisebox{6pt}[0pt]{\bf 290}
   &
   & \raisebox{6pt}[0pt]{$-0.6$} 
   & \\ 
 3 &  0.5 &  5.0 &  18.5 &  180 & 240 & 1270 & 17000 & {\bf 179} &      
   & 0.3 & $60-180$ \\ 
 4 &  1.1 &  6.2 &  15.3 &   25 & 40 &  &       &     &      
   &$-0.8$&          \\
 5 &      &      &       &  745 & 655 & 1600 &       &     & 0.25
   &      &           \\
 6 & \raisebox{6pt}[0pt]{0.2}
   & \raisebox{6pt}[0pt]{0.8}
   & \raisebox{6pt}[0pt]{24.4}
   & 355
   & 200
   & 915
   & \raisebox{6pt}[0pt]{25000}
   & \raisebox{6pt}[0pt]{249}
   &
   & \raisebox{6pt}[0pt]{1.1} 
   & \raisebox{6pt}[0pt]{$370-1350$}\\
 7 &      &       &  1.1  &   20 & & & 27000 & 354 & & 0.9
   & $10-600$ \\  
 8 &      &       &       &  230 & 300 &  &    &   &    &     &   \\
 9 &$<0.8$&   0.5 &  10.3 &  475 & 445 & 1430 & 15000 & {\bf 164} &      
   & 1.2 &$100-755$ \\
10 &  2.3 &  19.4 &  71.6 &  990 & 1495 & 8830 & 19000 & 360 & 0.59
   & 0.7 & 115  \\
11 & 28.2 & 134.8 & 347.4 & 860 & 2090 & 2260 & & {\bf 228} &      
   &$-0.1$ &$90-260$ \\
12 &  0.6 &   2.1 &   6.5 &   85 &  &    &       & {\bf 146} &      
   & 0.6 &     \\
\hline
13 & 64.9 & 109.3 & 124.2 &  245 &  &    &       &     &      
   &$-0.6$&        \\
14 &  0.3 &   1.7 &   8.6 &   30 & & $<2665$& 18000 & 180 & 2.84 
   &$-0.3$ &$80-995$ \\
15 &  1.7 &  14.6 &  46.6 &   75 & & $<2665$&       &     &      
   &$-0.9$&        \\
16 &  2.2 &   9.6 &  23.5 &   20 & & $<2450$&       &     &      
   &$-1.3$&        \\
17 & 29.8 &  37.6 &  58.4 &  140 &  &   790 &       &     & 0.24 
   &$-0.5$&        \\
18 &  0.4 &   2.2 &  8.0  &   20 &   &      &       &     &      
   &  0.2 &        \\
19 &      &       &       &   11 &   &   & 17000 & 340 &      
   &      & $10-600$  \\
20 &  1.3 &   6.8 &  23.8 & 1510 & 2275 & 3430 &       & {\bf 351} &      
   & 1.5  &$30-310$ \\
21 &      &       &       &  315 & 364 &  8665 &       &     &      
   &      &       \\
22 & \raisebox{6pt}[0pt]{9.3}
   & \raisebox{6pt}[0pt]{68.4}
   & \raisebox{6pt}[0pt]{189.2}
   & 1130
   & 1130
   & 1980
   & \raisebox{6pt}[0pt]{36000}
   & \raisebox{6pt}[0pt]{676}
   &
   & \raisebox{6pt}[0pt]{0.0}
   & \raisebox{6pt}[0pt]{$80-645$}\\
23 & 30.0 &  67.8 &  83.5 &  140 & 75 & $<2830$&       &     &      
   &$-0.7$&        \\
24 &  1.3 &   3.0 &   5.9 &   45 & & $<2830$& 67000 & 1252& 0.64 
   & 0.2 &$15-340$\\
25 &$<$0.6&   1.1 &   5.2 &   26 &   &      &       & 216 && 0.0 & \\ 
26 & 52.0 &  91.8 & 115.2 &  332 &    &     &       & 423 && $-$0.3 & \\
27 &  4.9 &  46.2 & 230.0 & 8510 &   &      & 34000 & 452 &      
   & 1.1 & 160 \\
28 &  0.6 &  11.2 &  88.7 & 2995 &  & 18010 &       & {\bf 186} & 1.04 
   & 1.1 & 1075  \\
29 &$< 0.1$&  0.4 &   5.4 &$< 50$&  &       & 15000 & 300 &      
   &$<0.5$&$15-705$ \\
30 & 40\hspace*{7.2pt}
   & 80\hspace*{7.2pt}
   & 90\hspace*{7.2pt}
   & 60 & &   &  & {\bf 186} & 
   &$-1.4$&  \\
31 &       &       & 170\hspace*{7.2pt} & 430 & 365 &   &    &   &   
   &$-0.4$& \\
32 & 176.9
   & 781.7
   &1635
   &4060 
   &4845 
   &
   & \raisebox{6pt}[0pt]{12000}
   & \raisebox{6pt}[0pt]{314} 
   & 
   &$-0.6$
   &\raisebox{6pt}[0pt]{$165-2760$}\\
33 &  4.0 &  24.3 &  56.8 &   75 & 685 &      &       &     &     
   &$-1.0$&     \\
34 & 30\hspace*{7.2pt}
   & 80\hspace*{7.2pt}
   & 110\hspace*{7.2pt}
   & 550 & 92 &  &  &  & 
   & 0.0 &     \\
35 & 17.8 &  54.4 &  79.4 &   60 &  &    &       &     &     
   &$-1.3$&       \\
36 & 32.4 &  52.2 &  57.2 &   39 &  &    &       & {\bf 184} &     
   &$-1.4$&       \\
37 & 31.2 &  43.0 &  35.8 &$< 65$&  &    & 15000 & {\bf 189} &     
   &$< -0.6$&$45-1845$ \\
38 & 12.1 &  24.4 &  25.2 &$< 55$&  &    & 15000 & {\bf 161} &     
   &$< -0.5$&$45-1890$ \\
39 &      &       &       &  255 &  &    &       &     & &&\\
40 & \raisebox{6pt}[0pt]{9.7}
   & \raisebox{6pt}[0pt]{31.5}
   & \raisebox{6pt}[0pt]{56.1}
   & 150
   &
   &
   &
   & \raisebox{6pt}[0pt]{{\bf 140}} 
   & 
   & \raisebox{6pt}[0pt]{0.2} 
   & \raisebox{6pt}[0pt]{$765-2170$}\\ 
41 &  3.2 &   4.2 &  3.4 &$< 65$&  &     &       & {\bf 42}  &     
   &$< 0.9$&  \\
42 &$< 0.1$&  0.7 &  8.9 &   80 &  &     & 20000 & {\bf 288} &     
   & 0.2 &$30-1445$ \\
43 &      &       &      &   44 &  &     &       &     &     &&\\
44 & \raisebox{6pt}[0pt]{24.9}
   & \raisebox{6pt}[0pt]{51.6}
   & \raisebox{6pt}[0pt]{65.6}
   & 74
   &
   &
   &
   & \raisebox{6pt}[0pt]{{\bf 48}} 
   & 
   & \raisebox{6pt}[0pt]{$-0.7$} 
   & \raisebox{6pt}[0pt]{}\\ 
45 & 174.5 & 368.0 & 418.5 & 30 &  &     &       &     &     
   &$-2.8$&     \\
\hline
\end{tabular}
\end{table*}

Despite this wealth of data, the evolutionary stage of many objects
is still unclear; this is true for the pure (sub)mm sources
as well as for some NIR sources with (sub)mm and radio counterparts.
A major reason for the uncertainty is the relatively large positional
offset between data at different wavelengths which makes a unique
correlation sometimes impossible. Another difficulty arises from the
classification of the objects in the absence of sufficient spectral
coverage -- particularly at shorter wavelengths. As suggested by
Andr\'e et al.~(\cite{andre}), the ratio of bolometric to sub-mm
luminosity may give an indication of the youth of a protostar; they
argue that the bolometric luminosity is a measure of the protostellar
mass, while the sub-mm luminosity represents the mass of the cold,
circumstellar matter. The ratio of both quantities should therefore
reflect the evolutionary stage of the protostellar system in the sense
that the sub-mm luminosity must decrease with the ongoing accretion of
the surrounding material. However, in practice the ratio $L_{\rm
bol}/L_{\rm smm}$ is difficult -- in most cases even impossible -- to
determine, because these earliest stellar stages evolve deeply within
dust clouds so that NIR observations -- if available at all -- suffer
from heavy extinction. In addition, satellite (or airborne) FIR
observations of protostars have been limited so far by sensitivity and
spatial resolution so that $L_{\rm bol}$ remains a poorly constrained
quantity in most cases. Thus, it seems that further classification
criteria for protostellar sources would be highly desirable.

The advent of sensitive ground--based MIR imaging devices is likely to
improve the situation because they i) provide a sub-arcsec resolution,
ii) have a better sensitivity than {\em IRAS}, and iii) are barely
influenced by interstellar extinction. In addition, they fill an
important gap in existing spectral energy distributions (SEDs) -- right
between the NIR and (sub)mm regimes,  and thus seem to be ideally
suited to elucidate the evolutionary stage of protostellar candidates.

In the present paper, we report on an $N$ and $Q$ band survey of
OMC\,2/3 with TIMMI\,2 following up our previous investigations of the
youngest stars in Orion. Data from our new extended $1200\,\mu$m
continuum survey of the region with SIMBA as well as NIR data from the
2MASS database are also included.


\section{Observations}

The mid-infrared (MIR) observations were carried out in January 2001
and 2002 with TIMMI\,2 (Dietzsch \& Reimann~\cite{dietzsch};
Reimann et al.~\cite{reimann}) at the ESO\,3.6\,m telescope, La Silla,
Chile. TIMMI\,2 is a mid-infrared spectrograph and imager operated in
the M (5\,$\mu$m), N (10\,$\mu$m) and Q (20\,$\mu$m) atmospheric bandpasses
(c.f. \texttt{http://www.ls.eso.org/lasilla/Telescopes/360cat/ timmi/}). We
observed in the $N$- and $Q$-bands. The first run was carried out
using a set of filters centred at wavelengths of 11.9 ($N$) and
17.8$\,\mu$m ($Q$). The $N$-band filter was replaced with one centred
at 10.4$\,\mu$m for the second campaign, in which we re-measured nearly all
the previous detections. The average seeing was about 0\farcs7 being
close to diffraction limited imaging. In order to achieve a good
relative pointing accuracy, we
used the bright object MMS\,7 (Chini et al.~\cite{chini}) as a
positional standard from which we offset the telescope for each new
target position. This enabled us to correlate the MIR sources in a
unique way with detections from the VLA, the 2MASS data base, and
(sub)mm observations. Several suitable TIMMI\,2 standard stars were
observed for calibration purposes.

\begin{figure}
\centering
\resizebox{\hsize}{!}{\includegraphics{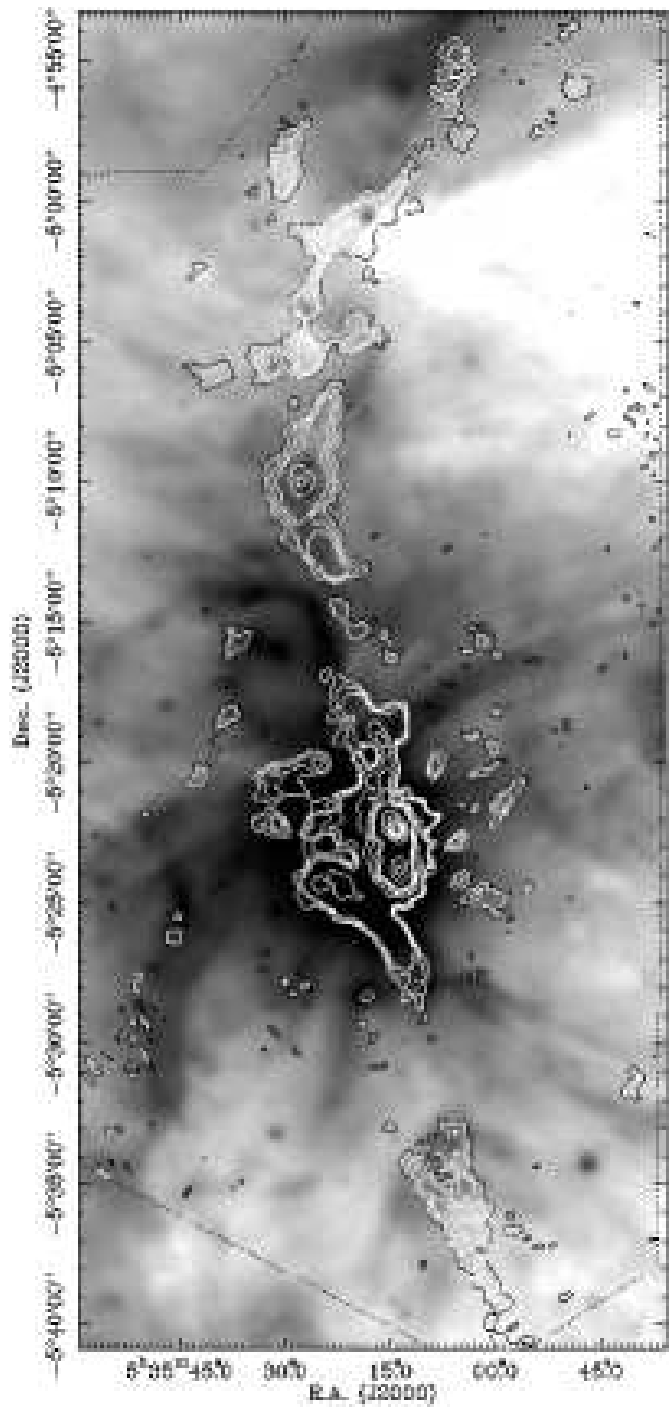}}
\caption{A SIMBA view of the Orion\,A region at $1200\,\mu$m
(contours) superimposed on an MSX image at $8.3\,\mu$m (grey-scale). In
the Northern region, dust filaments that are seen in emission with SIMBA
appear in absorption for MSX, indicating the presence of cool and
dense dust. The central part around the Kleinmann-Low nebula in OMC\,1 is
dominated by MIR emission suggesting considerably higher dust temperatures. }
\label{simbamap.fig}
\end{figure}

The $1200\,\mu$m continuum observations were carried out with the
37-channel bolometer array SIMBA at the SEST on La Silla, Chile during
the second commissioning period in October 2001. Skydips were
performed every three hours in order to correct for the atmospheric
opacity. Maps of Uranus were taken for calibration purposes. The final
map of about $1^\circ$ in declination covers the full region of OMC\,1
to 3 and was created from 13 single fast-scanning maps. The residual
noise is about 40\,mJy/beam (rms). The beam size is $24''$. All data
were reduced and analysed with MOPSI\footnote{MOPSI is a reduction
software for infrared, millimetre and radio data, developed and
constantly upgraded by R. Zylka, IRAM, Grenoble, France.}  according
to the instructions of the SEST manual~(\cite{sest}).

In addition, we show an extension of the spatially higher resolved
$1300\,\mu$m image of Chini et al.~(\cite{chini}) of 3$'$ towards the South.
This hitherto unpublished map was obtained at IRAM (Sievers,
priv. comm.) and allowed us to look for coinciding detections down to
a declination of $-5^\circ16'30''$.

The 2MASS data were accessed via the {\em VizieR Online Catalogue
Service}\footnote{\texttt{http://vizier.u-strasbg.fr}}. Only such
objects were selected which were detected in at least two wavebands of
$JHK_s$.  The cited $K_s$ magnitude was transformed into a $K$
magnitude according to Wainscoat \& Cowie~(\cite{wainscoat}).
Furthermore, the conversion from magnitudes to physical flux
quantities was achieved following the calibration information of the
2MASS Explanatory Supplement (Sect.~IV 5a).

\begin{figure}
\centering
\resizebox{0.91\hsize}{!}{\includegraphics{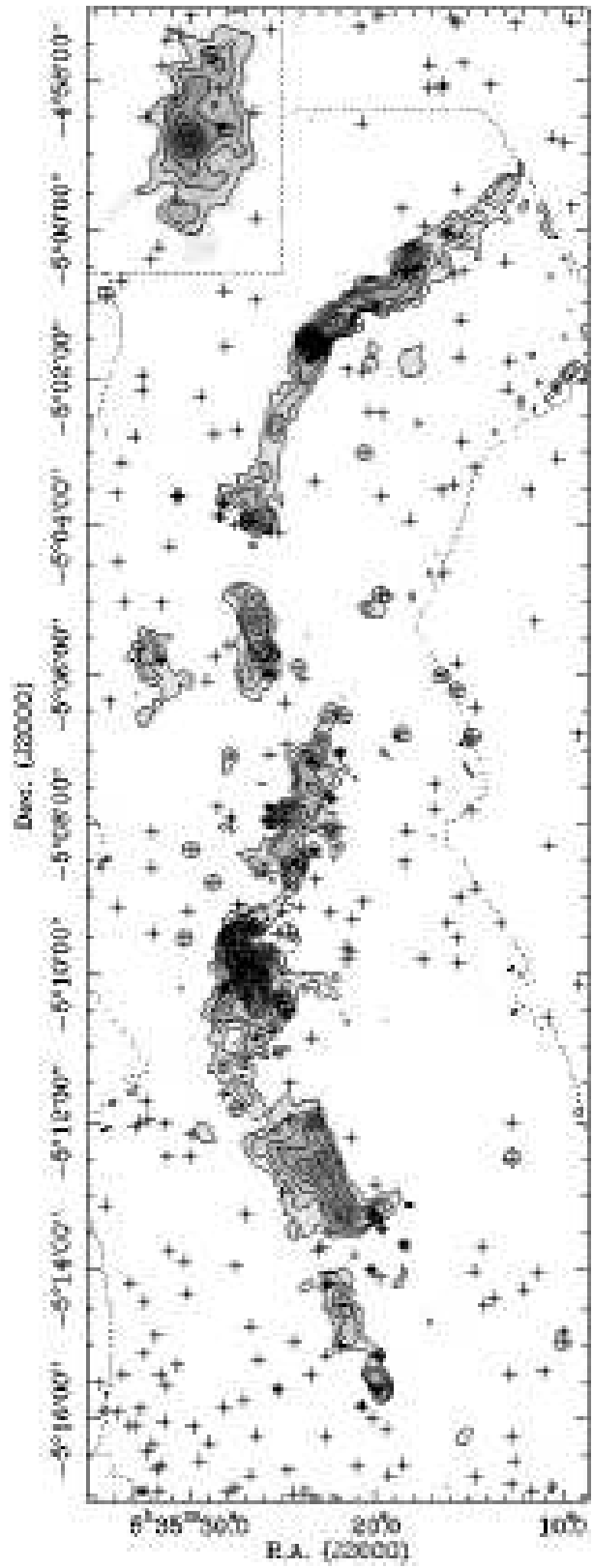}}
\caption{The OMC\,2/3 map at 1.3\,mm by Chini et al.~(\cite{chini})
is shown as a grey-scale enhanced with contours rising linearly from
a 3$\sigma$ r.m.s. noise level. Its Southern extension of $3'$ was
kindly provided by A. Sievers. The upper left insert was taken from
the SIMBA map. Our new MIR sources from TIMMI\,2 are marked as filled
dots while the 2MASS sources show up as crosses. Objects with a colour
excess are pointed out with an open circle.}
\label{iram.fig}
\end{figure}

\section{Overview of the region}

Fig.~\ref{simbamap.fig} gives an overview of the Orion~A region including
OMC\,1 -- 3; it was
obtained with SIMBA at $1200\,\mu$m and extends the original map by
Chini et al.~(\cite{chini}) both towards North and South. This map
resembles - apart from the lower resolution - very much our previous
one, but shows this time fainter extended emission. In addition, there
are new complexes with embedded sources in the Northern and the very
Southern region; the strongest compact source to the very North-East
is labelled ``S'' according to ``SIMBA''.  The $1200\,\mu$m contours
are plotted over an $8.3\,\mu$m image from MSX (Midcourse Space Experiment).
The correspondence at
both wavebands is striking. The $1200\,\mu$m emission filaments in
OMC\,3 appear in absorption in the MSX map, indicating the presence of
dense and cool dust. Moving further to the South, the dust temperature
seems to increase as witnessed by the emission at $8.3\,\mu$m.

Fig.~\ref{iram.fig} shows the $1300\,\mu$m emission of OMC\,2/3 at
twice the resolution of Fig.~\ref{simbamap.fig}; it was taken from
Chini et al.~(\cite{chini}) and was extended to the South by an
additional mosaic of $3'$ in declination, kindly provided by
A.~Sievers~(priv. comm.). The
positions of the new MIR sources from TIMMI\,2 as well as all 2MASS
detections within the region are also marked. It becomes clear from
this figure that we have to deal with a number of combinations as
concerns the coincidence of detections at NIR, MIR and mm wavelengths.

Table~\ref{pos.tab} gives the positions for all objects detected in
our MIR survey. Although there is a number of different names
associated with certain sources in the field, we use -- for reasons of
homogeneity - the running number of the MIR sources as a nomenclature
in the present paper. Other designations are given for comparison.
Comparing the positions derived  for our $10.4\,\mu$m
sources with astrometry from the VLA and 2MASS, we estimate an
accuracy of better than 1\arcsec. The VLA positions are taken from
Reipurth et al.~(\cite{reipurth}), while the NIR reference frame is
based on the 2MASS database. Whenever the the nominal MIR position
according to the telescope information agreed within 3$''$ with a
VLA or a 2MASS detection, we assumed that the
sources are identical. Since many MIR frames contain more than one
object and are often overlapping with adjacent frames, the
identifications could be cross-checked against each other and with
the MSX data, giving a consistent picture.

Altogether, there are 45 MIR sources which have 28 counterparts in the
$1300\,\mu$m map by Chini et al.~(\cite{chini}). However, only the 10
strongest mm sources were labelled with an MMS or FIR number by Chini
et al.~(\cite{chini}) while 17 weaker mm sources do not yet have any
millimetre designation. Those sources are marked with a ``+'' in
Table~\ref{pos.tab} in order to emphasise the coincidence between a
MIR and a $1300\,\mu$m source.

The NIR identifications in Table~\ref{pos.tab} refer to scan No. 10 of
the 2MASS survey as well as the measurements of Ali \&
DePoy~(\cite{ali}, AD), Carpenter et al.~(\cite{carpenter}, CHS) and
Gatley et al.~(\cite{gatley}), who introduced the labelling of IRS\,1
to 5. Additionally, those {\em IRAS} sources are mentioned, whose
error ellipses include the positions of our MIR sources.

The photometric data is given in Table~\ref{phot.tab}. It contains the
NIR flux densities, mostly taken from the 2MASS database. In rare
cases, where 2MASS data were not available, we used the measurements
of AD and CHS. The listed MIR flux densities were acquired with our
TIMMI\,2 observations. For reasons of homogeneity, we re-measured all
$1300\,\mu$m flux densities for the MIR sources. Therefore, the values
for those 15 objects which were not obviously outstanding in the
$1300\,\mu$m map by Chini et al.~(\cite{chini}) are new and are marked
in bold letters. The same holds for the new North-Eastern source,
where we extracted the millimetre flux from the SIMBA map. The
previously published values are in accord with our new estimates.


\section{Results}

The detection of an individual source at the different wavebands (see
Table~\ref{pos.tab}) gives a first indication that we are dealing with
a variety of sources at different evolutionary stages. Likewise,
Table~\ref{phot.tab} corroborates this view, because the relative
energy output at certain spectral regions changes tremendously from
source to source. In order to get a comprehensive view of the stellar
content in the region, we have used two major data sets: 43 sources
come from our MIR survey (hereafter MIR sample) whereas 264 sources
were obtained from the 2MASS database (NIR sample). Subsequently, both
groups were divided according to their location in OMC\,2 and 3,
respectively.

\begin{figure}
\centering \resizebox{\hsize}{!}{\includegraphics{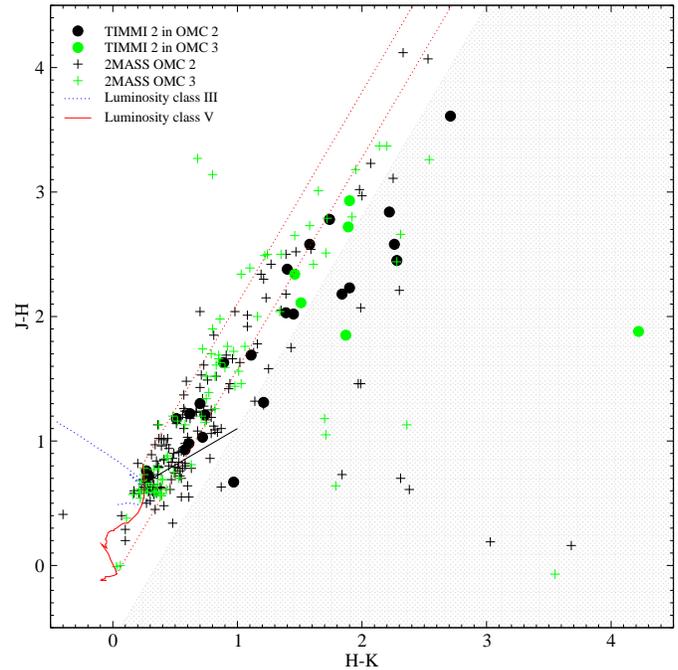}}
\caption{The NIR colour--colour diagram of 264 sources in the OMC\,2/3
region. The symbols follow our division into a MIR ($\bullet$) and a
NIR ($+$) sample; black symbols refer to sources in OMC\,2, grey
symbols to those in OMC\,3. The unreddened loci of MS stars (solid)
and giants (dotted) according to Ducati et al.~(\cite{ducati}) are
indicated as well as the locus of unreddened T\,Tauri stars, as
derived by Meyer et al.~(\cite{meyer}). The straight dotted lines denote the
direction of the normal reddening vector, the shaded area depicts the region
of NIR colour excesses.}
\label{jhk.fig}
\end{figure}

\begin{figure}
\centering \resizebox{\hsize}{!}{\includegraphics{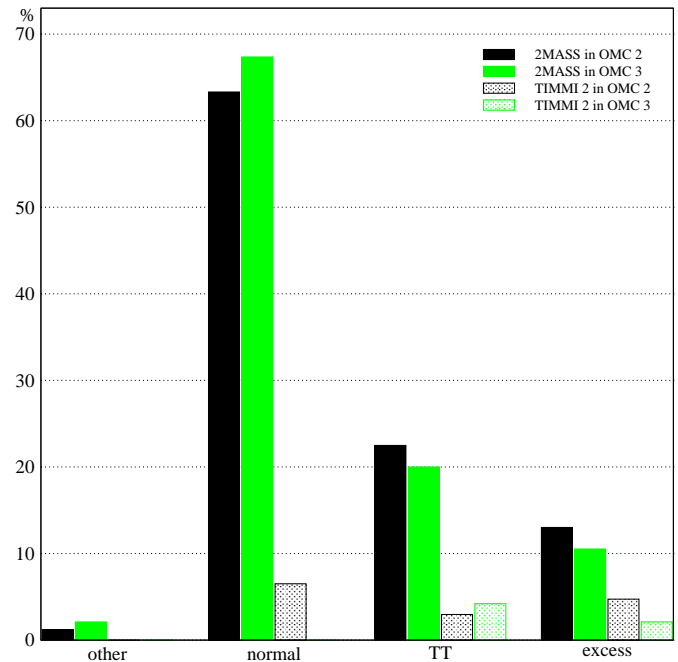}}
\caption{Statistics of all IR sources as they populate the different areas
of the colour--colour diagram. The fractions of sources relative to the
total number of IR sources separated in OMC~2 and OMC~3 are given.}
\label{excess_rel.fig}
\end{figure}

\subsection{The $JHK$ colour--colour diagram}

Fig.~\ref{jhk.fig} shows the NIR colour--colour diagram of all 264
known IR sources in the region; the photometric errors in both axes
are typically around 0.2~mag, and rise up to 1~mag for the faintest objects.
Despite this large error margin, the sources
are mainly located in ``allowed'' regions as discussed below.
Unfortunately, not all MIR sources can be associated with NIR
counterparts: 5 MIR binary systems have not been resolved by 2MASS and
thus have a single NIR entry only. 2 MIR objects were exclusively
detected at $K_s$ and 2 MIR objects could not be observed at all. In
order to achieve a crude classification of the sources we distinguish
four different areas within the $JHK$ diagram.

\begin{table}[h]
\caption{Statistics of sources in the four colour regions.}
\label{excess.tab}
\begin{tabular}{rcccc}
\hline
\hline
                 & \multicolumn{4}{c}{Colour region}       \\
\multicolumn{1}{c}{Sample}
                 & other & normal & T~Tau & excess \\
\hline
2MASS / OMC~2   & 2     & 107    & 38    & 22     \\
2MASS / OMC~3   & 2     & 64     & 19    & 10     \\
TIMMI~2 / OMC~2 & 0     & 11     & 5     & 8      \\
TIMMI~2 / OMC~3 & 0     & 0      & 4     & 2      \\
\hline
\end{tabular}
\end{table}

\subsubsection{Giant region}

Within the errors, 4 NIR sources lie in the region of giant stars. 3
of them have photometry errors of $> 0.4$~mag which makes a secure
classification as giant stars rather uncertain. Since 2 sources have
extinction values $15 < A_V < 20\,$mag, we regard them as true
background objects not related to the stellar population of the
OMC\,2/3 region. These 4 objects were excluded from any further
analysis.

\subsubsection{Main-sequence region}

There are 172 sources with "normal" colours in the area of reddened
main-sequence (MS) stars with $A_V$-values of up to 30~mag. Following
our crude division, the majority of objects ($161 \equiv 61$\%) comes
from the NIR sample only, indicating that this population is dominated
by evolved objects. On the other hand, only 11 objects (37\%) of the
MIR sample have NIR colours that are compatible with MS stars.

\subsubsection{T\,Tauri region}

Taking -- somewhat arbitrarily -- the lower reddening vector for MS
stars as the dividing line between MS and T\,Tauri (TT) stars we end
up with 57 sources that are located in the T\,Tauri region. They
exhibit $A_V$-values of up to 30~mag. Here, 48 objects (18\%) come
from the NIR sample alone. 9 objects of the MIR selected sample (30\%)
are found in that region. This suggests that the MIR group contains a
larger fraction of T\,Tauri stars.

\begin{figure}
\centering
\resizebox{\hsize}{!}{\includegraphics{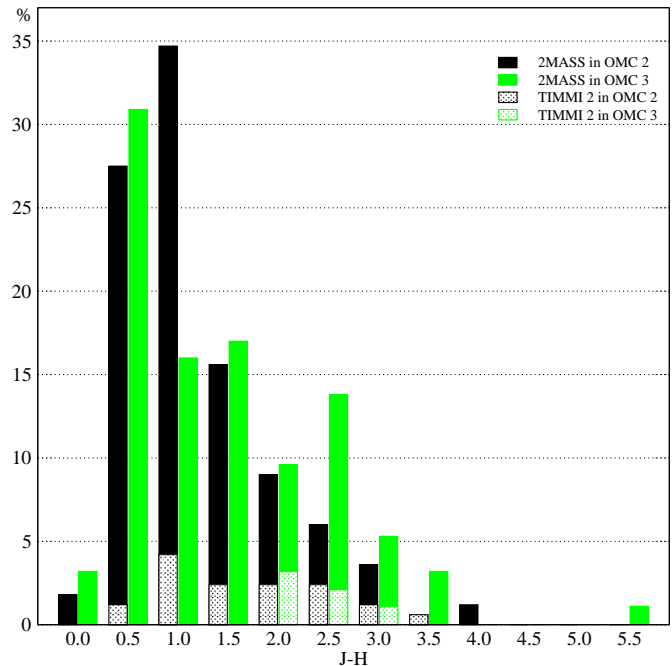}}
\caption{Relative distribution of $J-H$ colours normalised to the
number of 2MASS sources in OMC\,2 and 3, respectively.}
\label{colour}
\end{figure}

\subsubsection{Excess region}

The shaded area in Fig.~\ref{jhk.fig} denotes the regime of IR-excess
stars. Its left boundary to the T\,Tauri regime is determined by the
truncation of the T\,Tauri branch at $H-K = 1$, as shown by Meyer et
al.~(\cite{meyer}). Ten sources were detected at MIR wavelengths
(33\%), whereas 32 (12\%) could be found in the NIR. This
corroborates the trend that the fraction of YSOs increases in
the MIR sample compared to the NIR sample when going to earlier
evolutionary stages.

Altogether there are 32 excess objects of which 22 are located in
OMC\,2 and 10 in OMC\,3.  There is no clear division of both groups in
terms of extinction and/or strength of excess, although sources in
OMC\,3 seem to be slightly redder. For example, the source with the
highest extinction is the MIR~1/2 binary with $J-H > 5$, but it has
been excluded from the colour--colour diagram due to its large
error. The largest excess of $H-K > 4$ also occurs with an OMC\,3 object.

\begin{figure}
\centering \resizebox{\hsize}{!}{\includegraphics{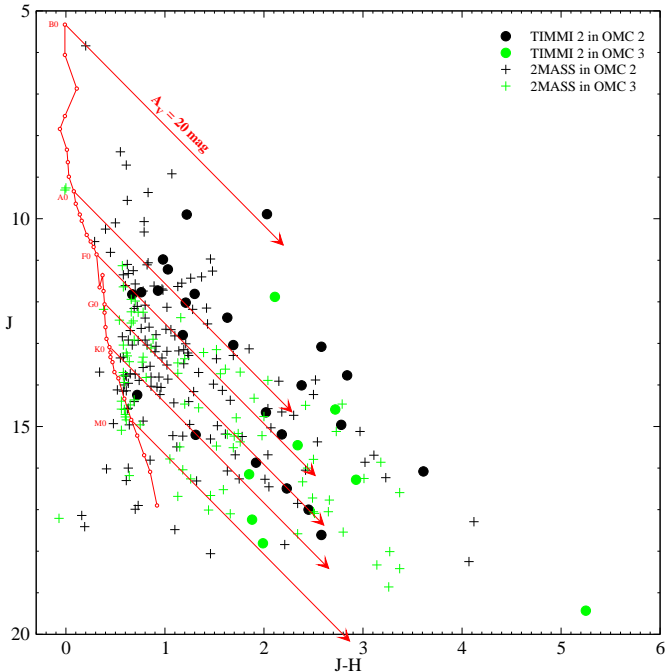}}
\caption{The NIR colour--magnitude diagram of 260 sources in the
OMC\,2/3 region, excluding the 4 sources lying in the area of
giants.  The symbols are as in Fig.~\ref{jhk.fig}. The MS has
been adjusted for a uniform foreground reddening of $A_V = 2$\,mag and
a distance of 450\,pc. Reddening vectors of a length corresponding to
$A_V = 20$\,mag were attached to the loci of unreddened types B0
to M0.}
\label{jh.fig}
\end{figure}

\subsection{The $J$ vs. $J-H$ colour--magnitude diagram}

Fig.~\ref{jh.fig} shows the $J$ vs. $J-H$ colour--magnitude diagram of
260 of the 264 sources excluding the 4 objects populating the area of
giant stars in Fig.~\ref{jhk.fig}; the symbols are the same as in
Fig.~\ref{jhk.fig}. The vertical curve represents the location of MS
stars from spectral types B0 to M4 (Schmidt-Kaler \cite{sk}; Ducati et
al.~\cite{ducati}), corrected for a foreground extinction of $A_V =
2$\,mag and a distance of 450\,pc; the arrows correspond to reddening
vectors of $A_V = 20$\,mag.  While a few objects experience an
extinction of up to $A_V=45$\,mag, most of them have $5 < A_V < 25$
without any obvious distinction between OMC\,2 and 3, as visible from
Fig.~\ref{colour}. Within the errors, all $J-H$ colours except $J-H=1.0$
and $2.5$ seem to be equally populated between OMC\,2 and 3 with a slight
trend to a stronger reddening in OMC\,3. However, we do not think that
these outliers justify the interpretation of a strong
indication for OMC\,3 sources to have redder colours than OMC\,2 sources.

As witnessed from the colour--colour diagram in Fig.~\ref{jhk.fig}, a number of
sources exhibits strong IR-excesses. Assuming that this effect mainly
dominates the $H-K$ colours, Fig.~\ref{jh.fig} should minimise the
influence of the excesses and thus yield reliable results.

All sources show a wide spread in magnitude and colour. The truncation
at $J \sim 18$\,mag is due to the detection limit of the 2MASS
observations. Likewise, the photometric errors increase with fainter
magnitudes. It is clear that the assignation of spectral types and
intrinsic extinction from a colour--magnitude diagram assumes that
there are neither foreground nor pre-main sequence stars; both would
simulate earlier spectral types. Background stars -- in contrast --
appear as later types. Unfortunately, neither uncertainty can be
removed from the present data set. 

\begin{figure}
\centering
\resizebox{\hsize}{!}{\includegraphics{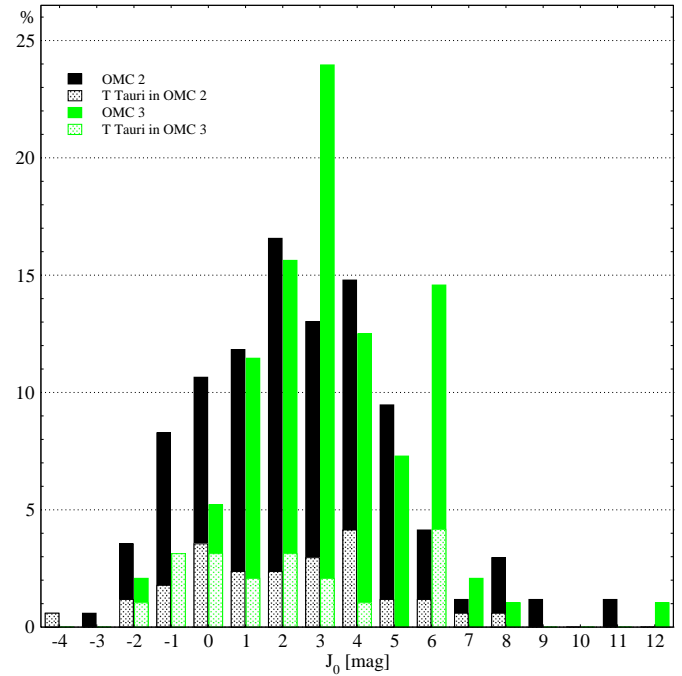}}
\caption{Relative distribution of $J$ luminosities normalised to the
number of 2MASS sources in OMC\,2 and 3, respectively.}
\label{luminosity}
\end{figure}

In order to minimise these uncertainties, we do not create a
distribution of spectral types, but construct a $J$ luminosity
function of all stars from Fig.~\ref{jh.fig} by de-reddening the
objects and correct for a distance of 450\,pc. This analysis should be
fairly independent of any colour excess, however it assumes that the
stars are on or close to the main sequence. In case of pre-main
sequence objects, the luminosity will be overestimated. On the other
hand, the commonly used $K$ luminosity function always simulates
higher luminosities, because it may be influenced by both an excess
caused by circumstellar emission and by pre-main sequence evolution;
both effects may increase the observed $K$ luminosity considerably.
Fig.~\ref{luminosity} displays the fraction of objects in a certain
$J$ luminosity interval relative to the  total number of 2MASS sources
in OMC\,2 and 3, respectively. The populations in both regions are
distributed across a luminosity range of $-4 < M_J < 12$ with a
maximum between $2 < M_J < 3$. The decrease of sources beyond this
maximum is certainly due to sensitivity. Nevertheless, there seems to
be a systematic shift between the two regions in the sense that OMC\,3
contains sources of lower luminosity. Neglecting the pre-main sequence
objects from the T\,Tauri region (dotted), the difference becomes even
more pronounced: the fraction of high luminosity, early type stars is
significantly higher in OMC\,2. Taking into account that OMC\,1 -- the
most Southern region of the Orion\,A complex -- has a rich abundance of
massive stars, it seems that there is a decrease of stellar masses
along the filament when going from South to North.

\section{Discussion}

In the following we discuss some particular sources of interest either
because of their morphology, multiplicity or their detection at other
wavelengths as well as the global properties of the association.

\subsection{Individual sources}

\subsubsection{OMC\,3~SIMBA}

\begin{figure}
\centering
\resizebox{\hsize}{!}{\includegraphics{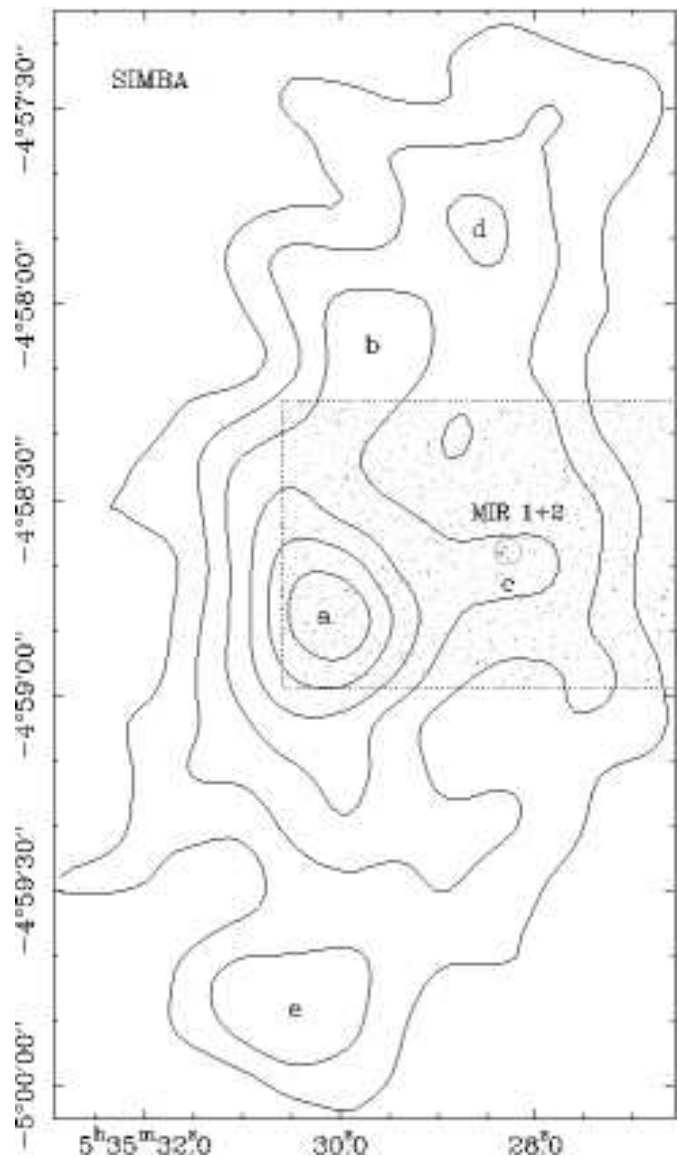}}
\caption{Superposition of the TIMMI\,2 image (rectangular grey-scale
region) and the SIMBA image (contours) of the region around
OMC\,3~SIMBA. The TIMMI\,2 image shows a binary source (MIR\,1/2) that
coincides with an extension from the main millimetre source, here
labelled as ``c''.}
\label{simba.fig}
\end{figure}

The new, North-Eastern complex (OMC\,3~SIMBA) has also been observed
by Lis et al.~(\cite{lis}) and Johnstone \& Bally~(\cite{johnstone}).
Molecular line measurements (source 2 of Tatematsu et
al.~\cite{tatematsu}) allowed to determine its gas mass to
140~$M_\odot$. In order to determine the dust temperature, we compare
this CO result with the total flux density at $1200\,\mu$m of 3.4\,Jy
and adopt a mass absorption coefficient $\kappa_\nu$ of 0.03\,cm$^2$
per gram of dust. The resulting dust temperature is then 11\,K
suggesting that OMC\,3~SIMBA is a fairly cold region; this picture is
corroborated by the fact that the complex shows up in absorption in
the MSX images (see Fig.~\ref{simbamap.fig}).

\begin{table}[h]
\caption{Positions of millimetre condensations of OMC\,3~SIMBA.  The
flux values are the peak intensities of the source positions in the
map. The $1 \sigma$ error is estimated to 30\,mJy/beam.}
\label{simba.tab}
\begin{tabular}{lrrc}
\hline
\hline
   & \multicolumn{1}{c}{R.A. (J2000)} & \multicolumn{1}{c}{Dec. (J2000)} & $S$ [mJy/beam] \\
\hline
a  & $5^{\rm h}35^{\rm m}30\fs2$ & $-4^\circ58'48''$ & 570 \\
b  &                    $29\fs7$ &         $58'06''$ & 330 \\
c  &                    $28\fs2$ &         $58'40''$ & 290 \\
d  &                    $28\fs6$ &         $58'48''$ & 290 \\
e  &                    $30\fs6$ &         $59'49''$ & 270 \\
\hline
\end{tabular}
\end{table}

As marked in Fig.~\ref{simba.fig}, OMC\,3~SIMBA contains at least five
condensations, labelled {\em a -- e} (see also Table~\ref{simba.tab}). The
$350\,\mu$m map of Lis et al.~(\cite{lis}) only shows the brightest
component while the more sensitive SCUBA maps of Johnstone \&
Bally~(\cite{johnstone}) resolve the structure, too. Unfortunately,
the authors did not elaborate on single sources in their paper, but
rather concentrated on the general structure of the filament.
Assuming that the total $350\,\mu$m flux density of 15\,Jy originates
from component ``a'', and taking the peak intensity at $1200\,\mu$m of
570\,mJy/beam, we obtain a sub-millimetre luminosity of
0.3~$L_\odot$. The spectral slope appears to be flat with $\beta =
0.7$. Using the above derived dust temperature of 11\,K we derive a
gas mass of 24\,$M_\odot$. However, it is likely that much of the
underlying extended emission contributes to the peak flux which
overestimates the flux density of the compact component ``a''. Fitting
e.g. a 2-dimensional Gaussian with an axis ratio of $33''\times24''$
to component "a" we derive a peak intensity of 375\,mJy/beam and a
total flux of 540\,mJy.

As far as we can judge from the astrometric accuracy, there is a
mid-infrared binary source (MIR\,1/2) that coincides with
component ``c''. The two objects are of equal strength at
$10.4\,\mu$m; their angular separation is 1\farcs2, and the position
angle is 30$^\circ$, measured clockwise from North. Using the
combined NIR colours from 2MASS, this binary system is compatible with
spectral type B.

\begin{figure}
\centering
\resizebox{\hsize}{!}{\includegraphics{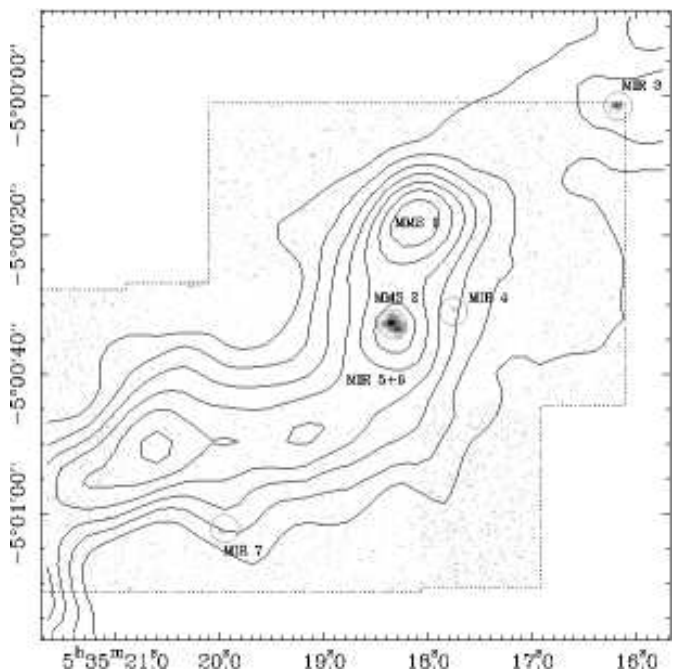}}
\caption{Superposition of the MIR and mm data of the region around
OMC\,3~MMS\,2 (MIR\,5/6). The grey-scale picture is the TIMMI\,2 image
showing three single and one binary source. The contours are taken
from the millimetre map of Chini et al.~(\cite{chini}).}
\label{mms2.fig}
\end{figure}

\subsubsection{OMC\,3~MMS\,2}

The millimetre source OMC\,3~MMS\,2 contains another MIR binary system
(MIR\,5/6) as revealed by our TIMMI\,2 imaging (see
Fig.~\ref{mms2.fig}).  The projected spatial separation of the
components is 1\farcs3 or 570\,AU with a position angle of
$45^\circ$. Both components of this binary system are surrounded by
extended emission at $10.4\,\mu$m. Subtracting the stellar point
sources, the residual emission suggests the presence of two
separate circumstellar discs. Tsujimoto et al.~(\cite{tsujimoto02})
resolved this binary with $JHKL$ photometry. They find a colour excess for
both sources corroborating the presence of circumstellar material. We will
propose high-resolution millimetre interferometry follow-up observations
in order to assess the actual distribution of the surrounding
material.

By applying Gaussians fits to the TIMMI\,2 sources,
we obtain similar ellipticities for both which would correspond to an
inclination angle of $33^\circ\pm2^\circ$ relative to the plane of the
sky. The combined NIR colours classify this system as K -- the only
late type binary system found in our sample -- with the largest excess
of $H-K \sim 4$. Analysing the two single YSOs individually, Tsujimoto et
al.~(\cite{tsujimoto02}) find similar excess values for both. The
colours from this study point to spectral types of A or B. From the
spectral index $\alpha$, we conclude that both components are of Class\,I.

MMS\,2 coincides with the X-ray source 8 (Tsuboi et
al.~\cite{tsuboi01}) while MIR\,5 is obviously identical to the X-ray
source 8c. The X-ray spectrum also shows evidence for two different
sources, one of which has a soft the other one a hard spectrum.
Tsuboi et al.~(\cite{tsuboi01}) interpret sources with soft X-ray
emission as T\,Tauri stars whereas hard X-ray emission is
characteristic for outflows and/or jets. Williams et
al.~(\cite{williams}) found a very long CO outflow associated with
MMS\,2. 

Another MIR source (MIR\,7) cannot be related to any of the millimetre
clumps in that region but it coincides with the X-ray source 12 of Tsuboi et
al.~(\cite{tsuboi01}). Its $JHK$ colours classify the star as T\,Tauri
which is compatible with its location far above the MS in the $J$ vs. $J-H$
colour--magnitude diagram. The spectral index $\alpha$ is compatible
with Class\,I.

\begin{figure}
\centering
\resizebox{\hsize}{!}{\includegraphics{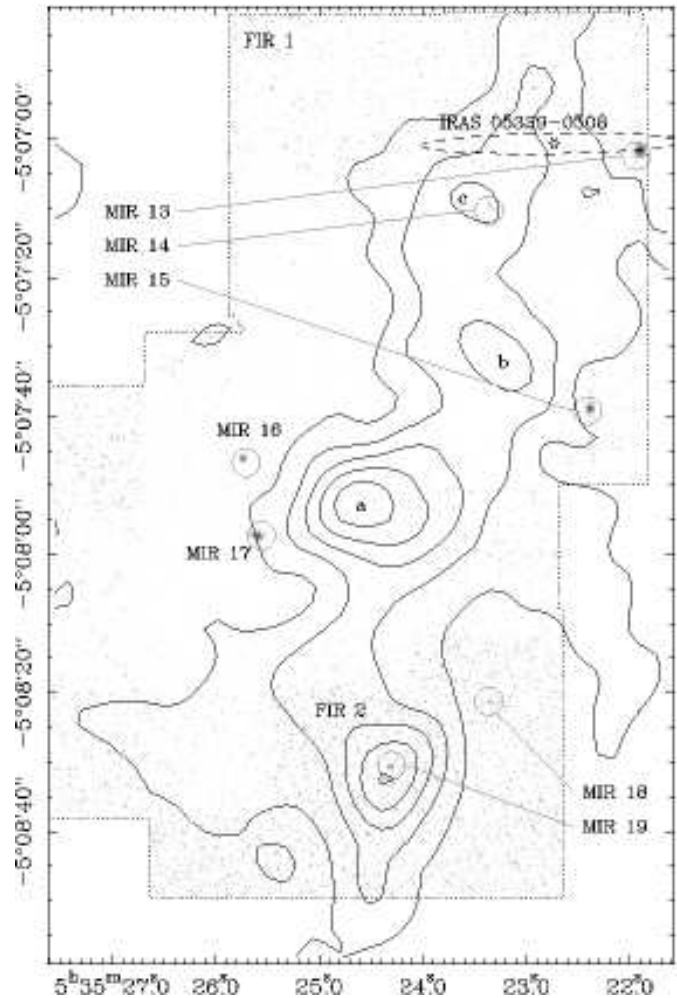}}
\caption{Superposition of the MIR and mm data of the region around
OMC\,2~FIR\,1 (MIR\,13--19). The grey-scale picture is the TIMMI\,2
image showing 7 sources. The contours are taken from the millimetre
map of Chini et al.~(\cite{chini}).}
\label{fir1f}
\end{figure}

\subsubsection{OMC\,2~FIR\,1--2}

The FIR\,1--2 complex harbours seven MIR sources, two of which (MIR\,14
and 19) are embedded in distinct $1200\,\mu$m condensations. MIR\,14
has the strongest excess in OMC\,2 with a $\Delta K$ of $\sim 1\,$mag
and an $M_J$ of 3.2; from its spectral index it is classified as
Class\,II. MIR\,19 was not detected by 2MASS. 

\begin{figure}
\centering
\resizebox{\hsize}{!}{\includegraphics{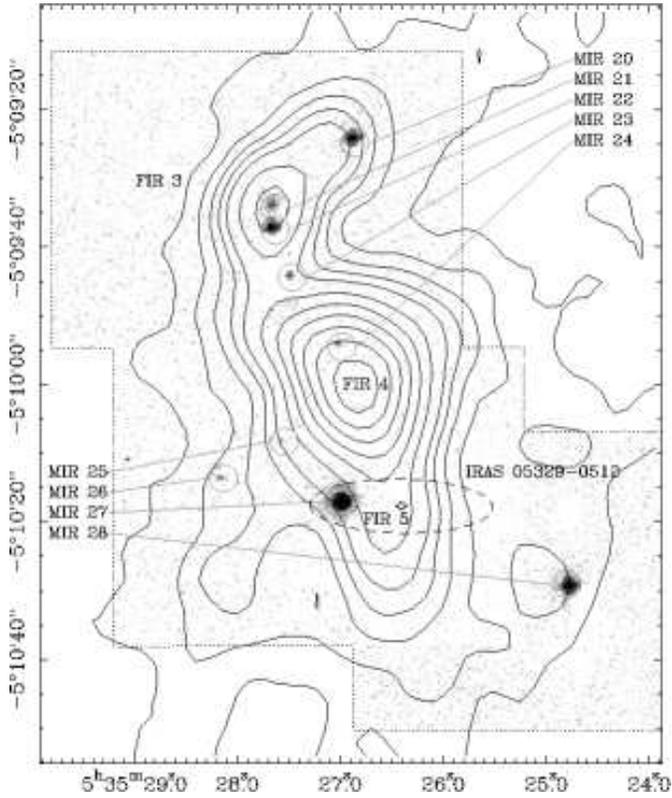}}
\caption{Superposition of the MIR and mm data of the region around
OMC\,2~FIR\,4 (MIR\,20--28). The grey-scale picture is the TIMMI\,2
image showing 9 sources. The contours are taken from the millimetre
map of Chini et al.~(\cite{chini}).}
\label{fir4.fig}
\end{figure}

\subsubsection{OMC\,2~FIR\,3--5}

Within the FIR\,3--5 complex we could detect nine MIR sources
(MIR\,20--28) of which five are associated more or less with
millimetre emission (see Fig.~\ref{fir4.fig}). MIR\,20 lies within an
unresolved North-Western extension of FIR\,3; its $JHK$ colours
display an excess of $\Delta K < 1$ while its spectral index $\alpha$
points towards Class\,I.

FIR\,3 contains a binary source (MIR\,21/22) with a projected
distance of about 1400\,AU. It was originally discovered as IRS\,4 by
Gatley et al.~(\cite{gatley}) and does not display any significant NIR
excess. Its spectral index $\alpha$ puts it among the Class\,II
objects although this classification might be contaminated by the high
visual extinction of $A_V \sim 22\,$mag.

FIR\,4 is in close neighbourhood to MIR\,24 which is a Class\,I source
with a tiny $K$ excess.

FIR\,5 is associated with an IRAS source and close to MIR\,27. Its
spectral index $\alpha$ suggests Class\,I; this is corroborated by its
$JHK$ colours that imply a small excess of $\Delta K \sim 0.5$.

MIR\,28 is a deeply embedded source with a $K$ excess of about 0.5. It
is located in a region of enhanced millimetre emission which, however,
has no separate nomenclature (see Fig.~\ref{fir4.fig}); it is also of
Class\,I.

\begin{figure}
\centering
\resizebox{\hsize}{!}{\includegraphics{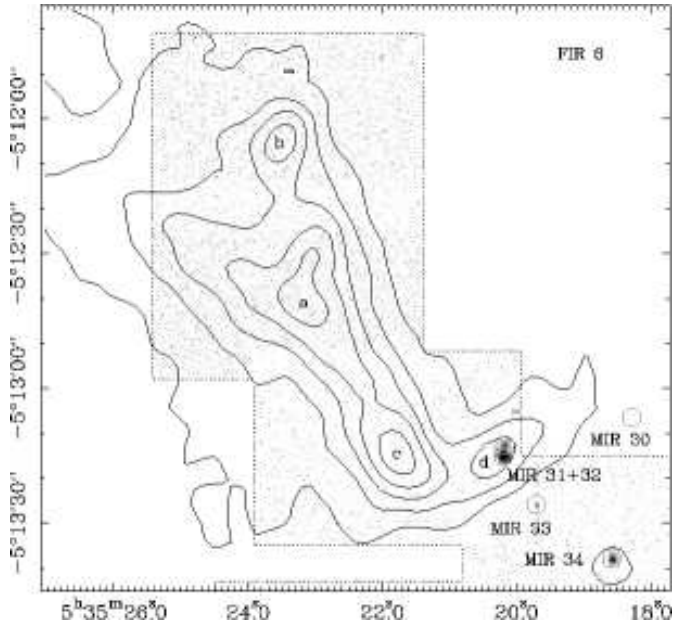}}
\caption{Superposition of the MIR and mm data of the region around
OMC\,2~FIR\,6 (MIR\,30--34). The grey-scale picture is the TIMMI\,2
image showing 5 sources. The position of MIR\,30 was determined from
the negative signal of the off-position. The contours are taken from
the millimetre map of Chini et al.~(\cite{chini}).}
\label{fir6.fig}
\end{figure}

\subsubsection{OMC\,2~FIR\,6}

Fig.~\ref{fir6.fig} displays FIR\,6 with its four millimetre clumps
{\em a -- d}. Only condensation {\em d} contains a binary MIR source (MIR\,31/32)
of which both components are of Class\,II; their projected distance is
1100\,AU. Another Class\,II source (MIR\,34) coincides with a faint
millimetre clump without nomenclature in the very South-Western part
of Fig.~\ref{fir6.fig}.

\subsubsection{OMC\,2~MIR\,35--45}

\begin{figure}
\centering
\resizebox{\hsize}{!}{\includegraphics{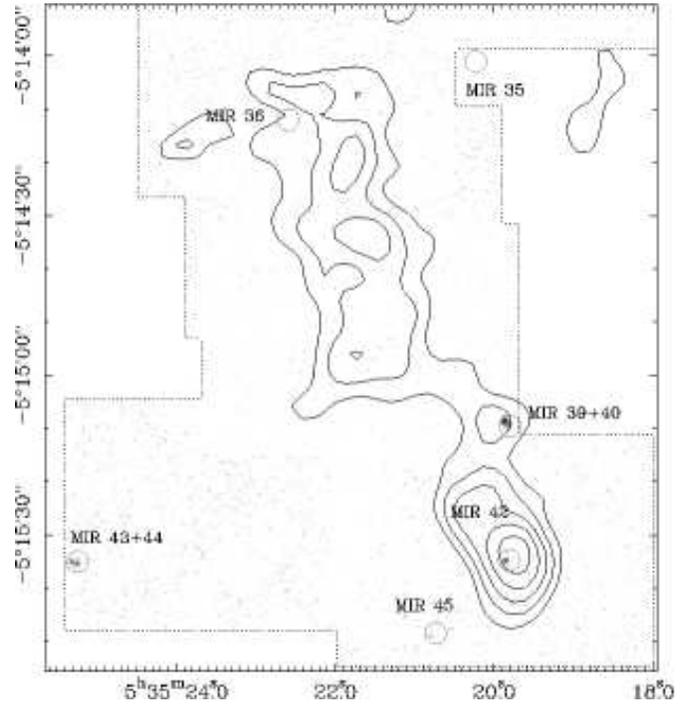}}
\caption{Superposition of the MIR and mm data of the region around
OMC\,2~MIR\,35--45. The grey-scale picture is the TIMMI\,2 image showing
8 sources. The position of MIR\,35 was determined from the negative signal
of the off-position. MIR\,37 and 38 only have upper limits. The contours
are taken from the Southern extension in Fig.~\ref{iram.fig}.}
\label{firsouth.fig}
\end{figure}

The new Southern extension of the millimetre map by Chini et
al.~(\cite{chini}) (see Fig.~\ref{iram.fig}) is enlarged in
Fig.~\ref{firsouth.fig}; it does not have any
nomenclatures. There is a rather strong millimetre peak in the very
South which coincides with the Class\,I source MIR\,42. Another
fainter millimetre clump contains the binary source MIR\,39/40 with a
projected separation of 380\,AU. A similar close binary (MIR\,43/44)
of Class\,II and a separation of 420\,AU lies far off the millimetre
emission ridge and at the Eastern edge of Fig.~\ref{firsouth.fig}.

\subsection{Global properties of the region}

Here we summarise the major properties of the extended dust emission
and the stellar association in OMC\,2/3 as they emerge from the
present study. Our new millimetre observations with SIMBA plus the
Southern extension of our imaging at IRAM reveals faint extended
emission filaments and enlarges the number of bona-fide dust
condensations to 28. A comparison between the extended MIR emission as
obtained from the MSX survey and the millimetre emission  shows that
the dust emission filaments seen by SIMBA in the Northern (OMC\,3)
region appear as absorbing infrared dark filaments at the MSX
wavebands, indicating the presence of cool and dense dust. Using the method
described by Siebenmorgen \& Kr\"ugel~(\cite{sieben}), we find
optical depths at $8.3\,\mu$m that correspond to visual extinctions of up to
200\,mag in the densest regions of OMC\,3. However, the locations of the
largest optical depths are not necessarily correlated with the strength of
the millimetre emission. In numerous cases, they lie well away from the
millimetre peaks. In contrast to that, we find that millimetre emission
regions farther to the South like e.g. OMC\,1, are dominated by the
radiation at MIR wavelengths, suggesting considerably higher dust
temperatures and lower densities. Free-free emission might also contribute
to the mm detections that are located within the \ion{H}{ii} regions of
the Orion nebula (OMC\,1).

Our MIR survey along the dust filaments of OMC\,2/3 has revealed 45
sources. Twelve of them belong to binary systems with projected
separations between 380 and 1400\,AU. These sources can also be
identified in the 2MASS survey where they are, however, unresolved.
Their combined NIR colours indicate that five systems consist of
early spectral types (probably B) while one seems to have
later type stars (K). Interestingly, all binaries are associated with
millimetre emission which probably arises from circumstellar discs. As
mentioned above, in the case of MIR\,5/6 the discs are to be seen
directly at $10.4\,\mu$m. In OMC\,2, the associated millimetre emission
decreases from North (676\,mJy) to South (48\,mJy) maybe suggesting
a proceeding disruption of circumstellar discs when approaching the
adjacent Orion nebula.

Concerning the evolutionary stages of the objects there is no
quantitative relation between the strength of the millimetre emission
and the evolutionary stage of the objects as derived from the spectral
index $\alpha$ between $K$ and $N$. At most there is a slight
statistical bias in the sense that two thirds of the Class\,I sources
are associated with millimetre clumps while only one third of the
Class\,II sources do show significant millimetre emission.
The distribution of $J$ luminosities shows that the fraction of
high luminosity early type stars is significantly higher in OMC\,2
compared to OMC\,3. On an even larger scale, the luminosities and thus
the masses of stars in Orion\,A seem to decrease from South (OMC\,1)
to North (OMC\,3).

\begin{acknowledgements}
We whish to thank the TIMMI\,2 team and the telescope team of the ESO
3.6\,m for their excellent support during this first observing run
with the new instrument. It is a pleasure for R.C. to thank N.~Ageorges
for her help and assistance with the observations and R.~Siebenmorgen
for helpful discussions concerning the calibration. We thank H.~Zinnecker
for his valuable ideas. Likewise, we want to thank the referee E.~Bica for
his helpful remarks.

This publication made use of data products from the Two Micron All Sky
Survey, which is a joint project of the University of Massachusetts and the
Infrared Processing and Analysis Center/California Institute of Technology,
funded by the National Aeronautics and Space Administration and the National
Science Foundation.
This research made use of data products from the Midcourse Space 
Experiment. Processing of the data was funded by the Ballistic 
Missile Defense Organization with additional support from NASA 
Office of Space Science. This research has also made use of the 
NASA/IPAC Infrared Science Archive, which is operated by the 
Jet Propulsion Laboratory, California Institute of Technology, 
under contract with the National Aeronautics and Space 
Administration.
\end{acknowledgements}


\begin{thebibliography}{}

  \bibitem[1995]{ali}
   Ali, B., \& DePoy, D. L. 1995, \aj, 109, 709 (AD)

  \bibitem[1993]{andre}
   Andr\'e, P., Ward-Thompson, D., \& Barsony, M., 1993, \apj, 406, 122

  \bibitem[1987]{bally}
   Bally, J., Langer, W. D., Stark, A. A., \& Wilson, R. W.
   1987, \apjl, 312, L45

  \bibitem[2001]{carpenter}
   Carpenter, J. M., Hillenbrand, L. A., \& Skrutskie, M. F.
   2001, \aj, 121, 3160 (CHS)

  \bibitem[1995]{castets}
   Castets, A., \& Langer, W. D., 1995, \aap, 294, 835

  \bibitem[1997]{chini}
   Chini, R., Reipurth, B., Ward-Thompson, D., Bally, J., Nyman, L.-\AA,
   Sievers, A., \& Billawala, Y. 1997, \apj, 474, L135

  \bibitem[1998]{dietzsch}
   Dietzsch, E., \& Reimann, H. G. 1998, SPIE, 2482, 151

  \bibitem[2001]{ducati}
   Ducati, J. R., Bevilacqua, C. M., Rembold, S. B., \& Ribeiro, D.
   2001, \apj, 588, 309

  \bibitem[1974]{gatley}
   Gatley, I., Becklin, E. E., Mattews, K., Neugebauer, G., Penston, M. V.,
   \& Scoville, N. 1974, \apj, 191, L121

  \bibitem[1998]{hillenbrand}
   Hillenbrand, L. A., Strom, S. E., Calvet, N., Merrill, K. M., Gatley, I.,
   Makidon, R. B., Meyer, M. R., \& Skrutskie, M. F. 1998, \aj, 116, 1816

  \bibitem[1990]{johnson}
   Johnson, J. J., Gehrz, R. D., Jones, T. J., Hackwell, J. A., \& Grasdalen,
   G. L., 1990, \aj, 100, 518

  \bibitem[1999]{johnstone}
   Johnstone, D., \& Bally, J. 1999, \apj, 510, L29

  \bibitem[1994]{jones}
   Jones, T. J., Mergen, J., Odewahn, S., Gehrz, R. D., Gatley, I.,
   Merrill, K. M., Probst, R., \& Woodward, C. E. 1994, \aj, 107, 2120

  \bibitem[1977]{kutner}
   Kutner, M. L., Tucker, K. D., Chin, G., \& Thaddeus, P. 1977, \apj, 215, 521

  \bibitem[1998]{lis}
   Lis, D. C., Serabyn, E., Keene, J., Dowell, C. D., Benford, D. J.,
   Phillips, T. G., Hunter, T. R., \& Wang, N. 1998, \apj, 509, 299 

  \bibitem[1996]{mc}
   McCaughrean, M. J., \& O'Dell, C. R. 1996, \aj, 111, 1977

  \bibitem[1997]{meyer}
   Meyer, M. R., Calvet, N., \& Hillenbrand, L. 1997, \aj, 114, 288

  \bibitem[1990]{mezger}
   Mezger, P. G., Wink, J. E., \& Zylka, R. 1990, \aap, 228, 95

  \bibitem[2001]{nyman}
   Nyman, L.-\AA., Lerner, M., Nielbock, M., Anciaux, M., Brooks, K.,
   Chini, R., Albrecht, M., Lemke, R., Kreysa, E., Zylka, R.,
   Johansson, L. E. B., Bronfman, L., Kontinen, S., Linz, H.,
   \& Stecklum, B. 2001, Msngr, 106, 40

  \bibitem[1993]{odell}
   O'Dell, C. R., Wen, Z., \& Hu, X. 1993, \apj, 410, 696

  \bibitem[1989]{rayner}
   Rayner, J., McLean, I., McCaughrean, M., \& Aspin, C. 1989, \mnras, 241, 469

  \bibitem[1998]{reimann}
   Reimann, H. G., Weinert, U., \& Wagner, S. 1998, SPIE, 3354, 865 

  \bibitem[1999]{reipurth}
   Reipurth, B., Rodr\'{\i}guez, L. F., \& Chini, R. 1999, \aap, 352, L83

  \bibitem[1982]{sk}
   Schmidt-Kaler, T. 1982, in Landolt-B\"ornstein, eds. Schaifers \& Voigt,
   Vol. VI/2b, Chpt. 4.1.2, 15

  \bibitem[2002]{sest}
   SEST operating manual 2002, ``The SEST Handbook'', Version 4.3, Chpts.
   6 and 8

  \bibitem[2000]{sieben}
   Siebenmorgen, R., \& Kr\"ugel 2000, \aap, 364, 625

  \bibitem[1993]{tatematsu}
   Tatematsu, K., Umemoto, T., Kameya, O., Hirano, N., Hasegawa, T.,
   Hayashi, M., Iwata, T., Kaifu, N., Mikami, H., Murata, Y., Nakano, M.,
   Nakano, T., Ohashi, N., Sunada, K., Takaba, H., \& Yamamoto, S.
   1993, \apj, 404, 643

  \bibitem[2001]{tsuboi01}
   Tsuboi, Y., Koyama, K., Hamaguchi, K., Tatematsu, K., Sekimoto, Y.,
   Bally, J., \& Reipurth, B. 2001, \apj, 554, 734

  \bibitem[2002]{tsujimoto02}
   Tsujimoto, M., Koama, K., Tsuboi, Y., Chartas, G., Goto, M., Kobayashi,
   N., Terada, H., \& Tokunaga, A.T. 2002, \apj, 573, 270

  \bibitem[1997]{yu}
   Yu, K. C., Bally, J., \& Devine, D. 1997, \apjl, 485, L45

  \bibitem[1992]{wainscoat}
   Wainscoat, R. J., \& Cowie, L. L. 1992, \aj, 103, 332

  \bibitem[2003]{williams} 
   Williams, J. P., Plambeck, R. L., \& Heyer, M. H. 2003, astro-ph/0303443,
   ApJ preprint doi: 10.1086/375396

\end{thebibliography}
\end{document}